\documentclass[twocolumn,showpacs,prl,amsmath,amssymb]{revtex4}

\usepackage{graphicx}
\usepackage{dcolumn}
\usepackage{bm}
\begin{document}

\title{Influence of a Uniform Current on 
Collective Magnetization Dynamics in a Ferromagnetic Metal}

\author{J. Fern\'andez-Rossier$^{ 1,2} $, M. Braun $^{ 1,3} $ 
A.S. N\'u\~nez$^1$ and A. H. MacDonald$^1$}
\affiliation{$^1$Physics Department, University of Texas at Austin,
Austin, TX 78712. USA}
\affiliation{$^2$ Departamento de F{\'\i }sica Aplicada, Universidad de 
Alicante,
San Vicente del Raspeig 03690, Alicante, Spain}
\affiliation{$^3$Institut f\"ur Theoretische Festk\"orperphysik, Universit\"at Karlsruhe,
76128 Karlsruhe, Germany}

\date{\today}

\begin{abstract}

We discuss the influence of a uniform current, $\vec{j}\;$, on the
magnetization dynamics of a ferromagnetic metal.  We find that the magnon
energy $\epsilon(\vec{q})$ has a current-induced contribution proportional to  
$\vec{q}\cdot \vec{\cal J}$, where $\vec{\cal J}$ is the spin-current,  and
predict that collective dynamics will be more strongly damped at finite ${\vec
j}$. We obtain similar results for models with and without local moment
participation in the magnetic order.  For transition metal ferromagnets, we
estimate that the uniform magnetic state will be destabilized for  $j \gtrsim
10^{9} {\rm A}\;{\rm cm}^{-2}$. 
We discuss the relationship of this effect to
the spin-torque effects that alter magnetization dynamics in inhomogeneous
magnetic systems.  

\end{abstract}
\narrowtext
\maketitle
	

\section{Introduction} 

The strong and robust magnetotransport effects that occur in metallic
ferromagnets (anisotropic, tunnel, and giant magnetoresistance for example
\cite{Maekawa}) result from the sensitivity of magnetization orientation to
external fields, combined with the strong magnetization-orientation
dependent potentials felt by the current-carrying quasiparticles. This
fundamentally interesting class of effects has been exploited in information
storage technology for some time, and new variations continue to be discovered
and explored . Attention has turned more recently to a distinct class of
phenomena in which the relationship between quasiparticle and collective
properties is inverted, effects in which control of the quasiparticle state is
used to manipulate collective properties rather than {\em vice-versa}.  Of
particular importance is the theoretical prediction \cite{Slon,Berger} of
current induced magnetization switching and related {\em spin transfer} effects
in ferromagnetic multilayers. The conditions necessary to achieve observable
effects have been experimentally realized and the predictions of theory largely
confirmed by a number of groups \cite{Tsoi1,Tsoi2,Sun,SMT-exp,Chien,MSU} over the
past several years.

Current-induced switching is expected \cite{Slon,Berger,TransferTheory} to
occur in magnetically inhomogeneous systems containing two or more weakly
coupled magnetic layers.   The work presented in the present paper was motivated by a
related theoretical prediction of Bazaily, Jones, and Zhang (hereafter BJZ), who argued
that the energy functional of a uniform bulk half-metallic ferromagnet contains
a term linear in the current of the quasiparticles \cite{BJZ}, {\em i.e.} that
collective magnetic  properties can be influenced by currents even in a
homogeneous bulk ferromagnetic metal.  The current-induced term in the energy
functional identified by  BJZ implies an additional contribution to the
Landau-Lifshitz equations of motion and, in a quantum theory, to a change
proportional to $\vec{q}\cdot\vec{j}$ in the magnon energy $\epsilon(\vec{q})$.
(Here $\vec{q}$ is the magnon or spin-wave wavevector and $\vec{j}$ is the
current density in the ferromagnet.)   The BJZ theory predicts that a
sufficiently large current density will appreciably soften spin waves at finite
wavevectors and eventually lead to an instability of a uniform ferromagnet. The
current densities necessary to produce an instability  were estimated by BJZ to
be of order $10^{8}$ A cm$^{-2}$, roughly the same scale as the current
densities at which spin-transfer phenomena are realized
\cite{Tsoi1,Tsoi2,SMT-exp,Chien,MSU}, 
apparently suggesting to some that these two phenomena are deeply
related. 

In this paper we establish that modification of spin-wave dynamics by current
is a generic feature of all uniform bulk metallic ferromagnets, not restricted
to the half-metallic case considered by BJZ.  We find that, in the general
case, the extra term in the spin wave spectrum
\begin{equation}
\delta \epsilon(\vec{q}) \propto \vec{q}\cdot\vec{\cal J}
\label{main}
\end{equation}
 where $\vec{\cal J}$ is the {\em spin current}, {\em i.e.}, the current
 carried by the majority carriers minus the current carried by the minority
 carriers \cite{footnote}. In the half
metallic case $\vec{\cal J}=\vec{j}$, recovering the result of Reference
\onlinecite{BJZ}. 
For reasons that will become apparent later, we refer to the extra term in the
spin wave dispersion as the {\em spin wave Doppler shift}, although this
terminology ignores the role of underlying lattice as we shall explain.   We
also study the effect of a uniform current on spin-wave damping. The usual
Gilbert damping law $\gamma \propto \epsilon(\vec{q}=0)$,  has an additional
contribution proportional to the  spin-current density.  In our picture, a
uniform current modifies collective magnetization dynamics because it alters
the  distribution of quasiparticles in momentum space.  

Our paper is organized as follows. In Section II we present two general
qualitative arguments which partially justify Eq.(~\ref{main}), independent of
any detailed microscopic model. In Section III we substantiate the arguments
with a microscopic calculation of the spin wave spectrum for a ferromagnetic
(but not necessarily half-metallic) phase of a Hubbard model, including the
effect of the current.  We derive Eq.(~\ref{main}), and demonstrate explicitly
that when  generalized from the half-metallic case to the general case, the 
spin wave Doppler shift is proportional to the spin-current {\em not the  total
current}. The microscopic calculation of Section III uses an effective action
approach, which separates collective and  quasiparticle coordinates in a
natural way and is well suited to study their interplay. In Section IV we
specialize to the half-metallic case and  re-derive the results of reference
\cite{BJZ} for the case of an $s-d$ model ferromagnet. This serves the purpose
of establishing a clear formal connection between the derivations presented in
Sections II and III and the derivation presented by BJZ, which appear
superficially to be quite distinct. In Section V we discuss the effect of a
current on spin wave damping. We consider both damping due to the coupling of
spin waves with the quasiparticles and two magnon damping, which we argue is
enhanced  by the spin wave Doppler shift of Eq.(~\ref{main}).  In Section 
VI we discuss the relationship between the spin wave Doppler shift and 
spin-transfer in inhomogeneous ferromagnets.  Finally, in  Section VII
we summarize our main results and present our conclusions.  

\section{Qualitative Explanation of the Current-Induced Magnon Energy Shift}

The low energy collective dynamics of the magnetization orientation in a ferromagnet
is described by the Landau-Lifshitz equation:
\begin{equation}
\frac{\hbar d\vec{\Omega}(\vec{r},t)}{dt}= \vec{\Omega}\times
\left[\frac{\delta E\left(\vec{\Omega},\partial_i \Omega_j\right)}
{\delta\vec{\Omega}} + \alpha \hbar \frac{d\vec{\Omega}(\vec{r},t)}{dt} \right]
\label{LL}
\end{equation}
where $\vec{\Omega}(\vec{r},t)$ is an unimodular vector field  which describes
the orientation of the collective magnetization and
$E\left(\vec{\Omega},\partial_i \Omega_j\right)$ is an energy functional of
$\vec{\Omega}(\vec{r},t)$ and its derivatives.  The generic applicability of this 
equation follows from the collective nature of spin-dynamics in ferromagnets.
It can be derived from a number of different microscopic models
in a number of different ways.  In particular, this equation describes the low-energy long-wavelength
dynamics of the two models of metallic ferromagnetism that we consider in later sections.
Normally $E$ is minimized by
a collinear configurations  $\vec{\Omega}(\vec{r},t)=\vec{\Omega}_0$ along
some privileged {\em easy} direction. The 
 Landau-Lifshitz equations linearized around $\vec{\Omega}_0$ have solutions
 which describe distortions of the magnetization orientation that propagate like
 waves with wave  $\vec{q}$ and frequency $\omega(\vec{q})$ \cite{Kittel}.  In a 
quantum treatment, magnetization orientation fluctuations are quantized in units of 
$\epsilon(\vec{q})= \hbar \omega(\vec{q})$.

In a metallic ferromagnet, the quasiparticles occupy bands\cite{bandpicture} that
are energetically split by an effective Zeeman-coupling magnetic field 
oriented along the direction $\vec{\Omega}$.
Non-collinear configurations are penalized because band-electron kinetic energies are raised by 
an inhomogeneous effective field $\vec{\Omega}(\vec{r},t)$. 
The easy axis is determined by spin-orbit interactions of the band
electrons and by the magnetostatic energy, which because of its long range 
depends on the overall shape of
the sample. 
 
The dynamics generated by the first term in square brackets in Eq.
(~\ref{LL}) is energy conserving whereas the second term, proportional to the
dimensionless coefficient $\alpha$, transfers energy from the collective
coordinate to other degrees of freedom.  In a metallic ferromagnet, the
damping is partly due to the excitation of electron-hole pairs in response to the
temporal evolution of $\vec{\Omega}$.
It is clear, therefore, that there is an intimate relation between the dynamics
of the collective coordinate and the state of the quasiparticles. What's more, when 
current flows inside a ferromagnet, the momentum-space distribution functions 
that describe quasiparticle state occupation probabilities are altered.
It is natural, therefore, to expect that 
the dissipative dynamics of the collective magnetization
will be affected by current flow. In Ref. \onlinecite{BJZ} it was shown
that, in a half metallic ferromagnet modeled by a s-d model (a model with a single band coupled
by exchange interactions to local moments), the energy
functional $E$ has a term linearly
proportional to the quasiparticle current, $\vec{j}$.  In the following 
paragraphs we present three arguments
to support the idea that the spin wave spectrum of any metallic
ferromagnet is modified by a uniform current 
in a manner similar to that suggested by Eq.(~\ref{main}).

We start with the simplest case, a half-metallic ferromagnetic electron gas, in which the 
current effect can be understood simply in terms of Galilean invariance.  The current 
carrying state of this system is simply one in which the entire electronic systems moves
along with a drift velocity $\vec{v}_D$.  A spin wave excitation is one in which the 
magnetization orientation precesses around the easy axis with frequency $\omega(\vec{q})$:
\begin{equation}
\hat \Omega = (\epsilon \sin(\vec{q}\cdot{\vec r}- \omega_0(\vec{q}) t),
\epsilon \cos(\vec{q}\cdot{\vec r} - \omega_0(\vec{q}) t), 1 - \epsilon/2)
\end{equation}
In the lab frame, the system is seen as moving with velocity $\vec{v}_D$,
and carrying current $\vec{j} = -n e \vec{v}_D$.  The fixed position 
$\vec{r}_L$ in the lab frame, has position $\vec{r}_L - \vec{v}_D t$ in the moving 
frame.   The precession frequency seen at a fixed lab frame position is therefore 
Doppler shifted to $\omega_0(\vec{q}) + \vec{q} \cdot \vec{v}_D$.  

This simple effect is the essence of the 
spin-wave Doppler shift.  In terms of the current density the spin-wave Doppler 
shift in the magnon energy is $\hbar \vec{q} \cdot \vec{j}/en$.  Systems of practical interest are neither Galilean invariant
nor, with a few possible exceptions, half metallic, however so a more detailed analysis is required to 
determine how the spin-wave Doppler shift is manifested in real systems.

A second useful point of view follows from considering a
single-mode-approximation for the  quantum spin-wave energy
$\epsilon(\vec{q})=\hbar\omega(\vec{q})$.   Elementary magnon  excitations of a
ferromagnet reduce the total spin projection along the easy axis by one unit
and add crystal momentum $\hbar \vec{q}$.  A state with the correct quantum
numbers  can be generated starting from the ferromagnetic ground state (or from
a state that carries  a uniform current) $|\Psi_0\rangle$ by acting on it with
the `magnon creation operator' 
\begin{equation}
s_{-}(-\vec{q}) = \sum_{i=1,N} s_{-i} \exp(i\vec{q}\cdot\vec{r}),
\label{sminus}
\end{equation}
where $s_{-i}$ is the spin-lowering operator for the $i$-th particle.  
Two-particle Greens functions constructed from this operator have poles with 
large residues at magnon excitation energies.  The single-mode approximation
consists of  using $|\Psi(\vec{q})\rangle \equiv s_{-}(-\vec{q})
|\Psi_0\rangle$ as a variational wavefunction for the magnon state at
wavevector $\vec{q}$.  Given this approximation for the magnon state, its  
excitation energy
\begin{equation}
\epsilon(\vec{q}) \equiv \frac{\langle \Psi(\vec{q})| {\cal H} |\Psi(\vec{q})\rangle} 
{\langle \Psi(\vec{q})| \Psi(\vec{q})\rangle} - E_0
\label{smaenergyformal}
\end{equation} 
can be expressed in terms of the expectation value of a commutator between the
general many-particle Hamiltonian $\cal{H}$ and either magnon creation or
annihilation operators and simplified to the following form:  
\begin{eqnarray}
& &\epsilon(\vec{q}) = \frac{\hbar^2 q^2}{2 m} + \nonumber \\
&& \frac{\hbar \vec{q}}{m} \cdot 
\frac{\sum_{ij} \langle \Psi_0 | s_{+i} s_{-j} \exp[i\vec{q}\cdot(\vec{r}_j-\vec{r}_i)] \vec{p}_j |\Psi_0\rangle}
{\langle \Psi_0 | s_{+}(\vec{q}) s_{-}(\vec{-q}) |\Psi_0\rangle} 
\label{smaexplicit}
\end{eqnarray}
The second term on the right hand side of Eq.(~\ref{smaexplicit}) is the magnon
Doppler shift term. In this term $s_{\pm i}$ and $\vec{p}_i$ are the spin
raising and lowering and momentum operators for particle $i$.  The numerator
and denominator of this term are, in general, complex two-particle correlation
functions.  The correlation functions are simplified when the 
ferromagnetic state is approximated by a Slater determinant with definite
occupation numbers for both majority ($\uparrow$) and minority ($\downarrow$)
spin momentum states, {\it i.e.} by the electron gas Stoner model ferromagnetic
ground state.  Then to leading order in $\vec{q}$ we find that the magnon
Doppler shift has the value
\begin{equation}
\delta \epsilon(\vec{q}) = \frac{\hbar \vec{q}}{m} \cdot \frac{\vec{\cal J}}{n_{\uparrow}-n_{\downarrow}}.
\label{parabolicbands}
\end{equation}
Eq.~\ref{parabolicbands} is most easily obtained 
by writing the operators whose expectation values need
to be evaluated as a sum of one-body and two-body terms and then using standard second quantization
identities.  The most important conclusion suggested by this equation is that, at least 
for parabolic bands, in generalizing the magnon Doppler effect from half-metallic ferromagnets
to ferromagnets with states of both spins occupied, the current is replaced by the spin-current 
$\vec{\cal J}$, and the density by the spin-density.    

Finally, the same result can be derived by considering a variational wave function for
the spin-wave state of a ferromagnetic metal in which all quasiparticle states
that are singly occupied share a common spinor that describes long-wavelength
spatial precession around the easy direction.  For example if the  $\hat x$
direction is the easy direction the spinor that describes small amplitude 
precession is $(u,v)=(1-\eta^2/2,\eta \exp(i\vec{q}  \cdot \vec{r}))$.  The
$\vec{q} \cdot \vec{\cal J}$ correction  then follows by observing that the
magnon energy equals  the energy change divided by the change in the $\hat x$
direction magnetization component, with both quantities being proportional to
$\eta^2$ at small $\eta$.   This findings suggest that the explicit
approximate expression for the magnon Doppler shift, derived from the SMA for parabolic
bands, is likely to qualitatively correct even for realistic ferromagnets with 
more complicated band structures.  Indeed, that is the conclusion that follows 
from the more microscopic derivations in the following two sections. 

\section{Current driven spin waves in a Hubbard Model Ferromagnet}

In order to explain our theory of the influence of uniform currents on 
the spin-wave spectrum, we first recall how spin-waves and quasiparticle
states are related in equilibrium.  This development will also establish
the notation we use for the non-equilibrium case.  The description we use
is one in which a collective fluctuation field interacts with 
fermionic quasiparticle fields.  It allows us to borrow
from standard theories of quantum harmonic oscillators weakly coupled
to a bath, in order to generalize the theory of collective dynamics from
equilibrium to non-equilibrium cases.     

\subsection{Hamiltonian and effective action}

In the previous section we discussed three general arguments in support of the 
existence of a spin wave Doppler shift in a metallic ferromagnet that is
proportional to the spin current as in Eq.(~\ref{main}).   We now look more
closely at the underlying physics by carrying out an explicit microscopic
calculation of the spin waves for a Hubbard model in the presence of a current.
Unlike the s-d model considered in Ref. \onlinecite{BJZ}, the Hubbard model
allows for ferromagnetism in a system with only itinerant electrons. The 
Hubbard model
Hamiltonian is \cite{Moriya} :
\begin{equation}
{\cal H}=\sum_{i,j} t_{ij} c^{\dagger}_{i,\sigma}c_{j,\sigma}
+ U \sum_j n_{j,\uparrow}n_{j,\downarrow}
\end{equation}

The elementary excitations of a metallic ferromagnet are quasiparticles and
spin waves.  We want to derive the propagator for the spin waves of the
ferromagnetic phase of this model and to see how is affected by a quasiparticle
current.  To do so,  it is convenient to use the functional integration
approach, \cite{Prange4,Moriya,Hubbard,Schulz}, in which the quasiparticles are
integrated out and an effective action for the spin waves is obtained. This
procedure is sketched below, the details can be found in Refs.
\onlinecite{Prange4,Moriya,Hubbard,Schulz,Fradkin}.  The final result for the
spin wave spectrum is equivalent to that obtained by doing a Random Phase
Approximation (RPA) \cite{Doniach} calculation. However, the effective action
approach provides a convenient conceptual framework to understand the 
connection between spin waves and non-equilibrium quasiparticle states, the
central focus of this paper. 

The interaction term in the Hubbard model can be written as \cite{Moriya} 
$$U \sum_j n_{j,\uparrow}n_{j,\downarrow}=
 -\frac{2}{3} U \sum_{i} \vec{S}^2_i + \frac{U}{2}\sum_{\sigma,i}
n_{\sigma,i}$$
We represent  the partition function of this model as  a  path  integral over
fermion coherent states \cite{Negele}, labeled by
$\{\overline{\Psi}_{\alpha},\Psi_{\alpha}\}$ , where $\alpha\equiv{i,\sigma}$. 
The key idea which allows quasiparticle and collective degrees of freedom to be
separated, while still treating the magnetization as a quantum field, is the
introduction of a  Hubbard-Stratonovich transformation \cite{HS} to represent
the interaction term.  By making this transformation, we  trade a problem of
interacting fermions for a problem of independent fermions whose spin is
coupled to a bosonic spin-splitting effective magnetic field
$\vec{\Delta}_i(\tau)$, which acts as the collective magnetic coordinate.  The
partition function reads:
\begin{eqnarray}
{\cal Z}=\int {\cal D} \overline{\Psi}_{\alpha}(\tau) {\cal D} \Psi_{\alpha}(\tau) 
{\cal D} \vec{\Delta}_i(\tau)  
\exp\left[ -S\left(\overline{\Psi}_{\alpha},\Psi_{\alpha},\vec{\Delta}_i\right)
\right]
\nonumber
\end{eqnarray}
where the action is
\begin{eqnarray}
S
=\int_0^{\beta} d\tau 
 \sum_i \frac{3\vec{\Delta}_i(\tau)^2}{8 U} +
 \sum_{i,i',\sigma}
\overline{\Psi}_{i,\sigma}(\tau) {\cal G}^{-1}_{ij,\sigma,\sigma'}
 \Psi_{j,\sigma}(\tau) 
\label{action}
\end{eqnarray}
and
\begin{eqnarray}
{\cal G}^{-1}_{ij,\sigma,\sigma'}=\left( \partial_{\tau} -
\mu+\frac{U}{2}\right)\delta_{i,j}
+ t_{i,j} - \vec{\Delta}_i\cdot\frac{\vec{\tau}_{\sigma,\sigma'}}{2}
\delta_{i,j}
\label{G}
\end{eqnarray}
is the inverse of the Green's function operator.

The action (\ref{action}) is the sum of three terms, {\em i)} 
non-interacting tight binding fermions (with a Hartree shift), 
{\em ii)} a  term
quadratic in the bosonic field and 
{\em iii)} their coupling $\vec{\Delta}_i\cdot
\vec{S}_i$, where $\vec{S}_i(\tau)\equiv
\sum_{\sigma,\sigma'}\frac{1}{2}\overline{\Psi}_{i,\sigma}\vec{\tau}_{\sigma,\sigma'}
\Psi_{i',\sigma'}$. Since the action is quadratic in the fermion variables, 
the fermion functional integral can be formally evaluated. This allows to write the
partition function as a path integral over the auxiliary field $\vec{\Delta}$
only,

\begin{equation}
{\cal Z}=\int {\cal D} \vec{\Delta}_i(\tau)  
e^{ -S_{\rm eff}(\vec{\Delta})}
\label{z2}
\end{equation}

where the effective action reads:
\begin{equation}
S_{\rm eff}(\vec{\Delta})=
\int_0^{\beta} d\tau \sum_i \frac{3\vec{\Delta}_i(\tau)^2}{8 U}
- {\rm Tr}\; {\rm Ln} \left[ {\cal G}^{-1}(\vec{\Delta})\right] 
\label{seff}
\end{equation}
Eqs.(~\ref{z2}) and (~\ref{seff}) are one of the many possible
representations of the {\em exact} partition function for the Hubbard Model.
The effective action (\ref{seff}) describes a complicated quantum field theory
for  $\vec{\Delta}_i(\tau)$.   

\subsection{Mean Field theory: Spin-split bands}

The first step in a field theory of ferromagnetism is usually to look for classical solutions,
{\em i.e.} for field configuration $\vec{\Delta}^{\rm cl}_i(\tau)$ for which the
effective action is stationary. The saddle point equation reads 
$\vec{\Delta}^{\rm cl}_i= \frac{4U}{3}\langle \vec{S}_i \rangle$, where the
average is computed with a Green function ${\cal G}(\vec{\Delta}^{\rm cl})$
obtained  by replacing, in Eq.(~\ref{G}),  the fluctuating field
$\vec{\Delta}_i(\tau)$ by the saddle point solution.

Assuming the existence of a ferromagnetic mean-field state, the classical
solution  for a perfect crystal is static (independent of $\tau$) and
homogeneous (independent of $i$).  It is therefore  characterized by a
direction $\bf n$ and a length $|\vec{\Delta}^{\rm cl}|\equiv \Delta$. Because
of the spin rotational invariance of the Hubbard Hamiltonian, $\bf n$ is
arbitrary. In real systems $\bf n$ is determined by spin-orbit interactions and
magnetostatic effects. The mean-field Green's function,  ${\cal
G}(\vec{\Delta}^{\rm cl})$, describes fermions which occupy bands that are
spin-split by an effective magnetic field along $\bf n$ (See Fig.(1)). The
magnitude of the spin splitting, $\Delta$, is obtained from the saddle point
equations, which, for this simple model, reduce to the following form:
\begin{equation}
\Delta=\frac{4U}{3}\frac{1}{2{\cal N}}
\sum_{\vec{k}} \big[ n_F\left[\epsilon^{\uparrow}_{\vec{k}}\right] -
n_F\left[\epsilon^{\downarrow}_{\vec{k}}\right] \big]
\label{mf}
\end{equation}
where $\epsilon^{\sigma}_{\vec{k}}= \epsilon(\vec{k}) - \sigma \frac{\Delta}{2}$
are the quasiparticle energies of the spin-split bands and ${\cal N}$ is the number of lattice sites. 
Notice that the majority band has spins parallel to $\bf n$, 
denoted by $\uparrow$.
The saddle point equations show explicitly that the auxiliary
field $\vec{\Delta}^{\rm cl}$ is proportional to the average
fermion magnetization, which usually appears as the fundamental field in 
classical micromagnetic theories for realistic magnetic materials.
Hereafter we refer to $\vec{\Delta}(\tau)$ as the

\begin{figure}
\includegraphics[width=2.6in]{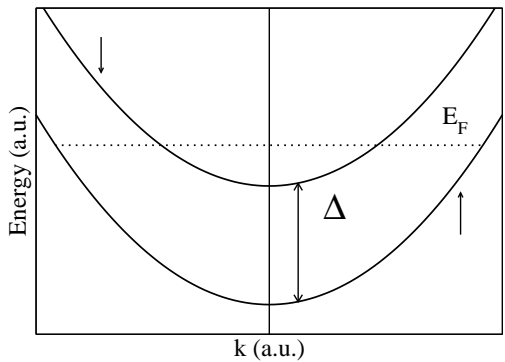}
\caption{ \label{fig1}Mean field quasiparticle bands. Dashed line shows the
Fermi Energy.  $\Delta$ is the spin splitting energy. }
\end{figure}

\subsection{Spin waves without current}

We are interested in the {\em dynamics} of  the collective coordinate, so that
the static solution obtained by solving the mean field approximation is insufficient.
To describe the elementary collective excitations 
we need to study small amplitude dynamic 
fluctuations of the collective coordinate around the static solution:
\begin{equation}
\vec{\Delta}_i(\tau)\simeq \vec{\Delta}^{\rm cl} + \delta\vec{\Delta}_i(\tau)
\label{fluc}
\end{equation}
We introduce Eq.(~\ref{fluc}) into the effective action (Eq.(~\ref{seff})) and 
neglect terms of order $\left[\delta\vec{\Delta}_i(\tau)\right]^3$ and higher. 
The resulting action$S_{\rm cl}(\vec{\Delta}^{\rm cl}) + S_{\rm SW}$, where the first term
is the classical approximation to the effective action and
the fluctuation correction is:
\begin{equation}
S_{\rm SW}= \frac{1}{2\beta \cal N} 
\sum_{{\cal Q}, a,b}
 \delta{\Delta}_a({\cal Q}) 
 {\cal K}_{ab}({\cal Q})
 \delta{\Delta}_b(-{\cal Q})
 \label{spinfluc}
\end{equation}
where ${\cal Q}$ is a shorthand for $\vec{q},i\nu_n$, and $a,b$ stand
for Cartesian coordinates. Note that the bosonic fields, 
$\delta\vec{ \Delta}(\cal Q)$ are dimensionless and the Kernel ${\cal K}$ has
dimensions of inverse energy.  This action defines a field
 theory for  the spin fluctuations. The equilibrium Matsubara Green function, 
${\cal D}_{ab}(\vec{q},i\nu_n)$ ,
is given  \cite{Auerbach,Negele} by the inverse of spin fluctuation Kernel, ${\cal K}_{ab}({\cal
Q})$. Analytical expressions for 
${\cal K}_{ab}({\cal Q})$ are readily evaluated for the case of parabolic bands 
and are appealed to below.
 We obtain the retarded spin fluctuation propagator by
 analytical continuation of the Matsubara propagator:
  $D_{ab}^{\rm ret}(\vec{q},\omega)
  ={\cal D}_{ab}(\vec{q},i\nu_n\rightarrow \omega+i0^+)$
 The imaginary part of the retarded propagator summarizes 
the spectrum and the damping of the spin fluctuations most directly. 
 
The theory defined by Eq.(~\ref{spinfluc}) includes two types of  spin
fluctuations which are very different: {\em i)} longitudinal fluctuations
(parallel to $\bf n$),   or amplitude modes and {\em ii)}  transverse
fluctuations (perpendicular to $\bf n$), or spin waves. The amplitude modes
involve a change in the magnitude of the local spin splitting, $\Delta$, and are either over damped 
or appear at energies above the continuum of spin-diagonal particle-hole excitations.
In contrast, the spin waves are gapless in the limit $\vec{q}=0$,
in agreement with the Goldstone theorem, and are often weakly damped even in 
realistic situations.  Note that the amplitude modes
decouple from the spin wave modes for small amplitude fluctuations. 
For $\hat x=\bf n$, we can write
\begin{equation}
{\cal K}_{ab}({\cal Q}) = \left[  
\begin{array}{ccc}
{\cal K}^{||} &0 &0 \\
0 &{\cal K}_{yy} &{\cal K}_{yz} \\
0 &{\cal K}_{zy} &{\cal K}_{zz} 
\end{array}
\right]
\end{equation} 
 Since the low energy dynamics of a metallic ferromagnet is governed by
transverse spin fluctuations, we do not discuss 
longitudinal fluctuations further.  After analytic continuation, we obtain
 the following result for the inverse of the retarded transverse spin
fluctuation Green function $(D^{\rm ret})^{-1}$, which is diagonal
 when we rotate from $\hat y,\hat z$ to
 $ +\hat z \pm i\hat{y}$ chiral representations.
 The diagonal elements are then
 \begin{eqnarray}
D^{\rm ret}_{\pm}(\vec{q},\omega)
&=& \frac{4U}{3}\frac{1}{1+ \frac{2}{3}U\Gamma(\pm\vec{q},\pm\omega)} 
\label{prop0}
\end{eqnarray}
where $\Gamma(\vec{q},\omega)$ is the Lindhard function evaluated with the spin-split
mean-field bands:
\begin{equation}
\Gamma(\vec{q},\omega)=\frac{1}{\cal N} \sum_{\vec{k}}
\frac{n^{\uparrow}_{\vec{k}}-
n^{\downarrow}_{\vec{k}+\vec{q}}}
{\epsilon^{\uparrow}_{\vec{k}}-\epsilon^{\downarrow}_{\vec{k}+\vec{q}}
 + \omega+i0^+}
\label{Linhardt}
\end{equation}
where $n^{\sigma}_{\vec{k}}$ is shorthand for the Fermi-Dirac occupation
function $n_F\left[\epsilon^{\sigma}_{\vec{k}}\right]$ for the quasiparticle
occupation numbers.
Eqs.  (~\ref{prop0}) and (~\ref{Linhardt}) make it clear that the spin wave
spectrum is a functional of the occupation function $n_F$ for the
quasi-particles in the spin-split bands.  The influence of a current on 
the spin-wave spectrum will enter our theory through non-equilibrium 
values of these occupation numbers. 

In the case of parabolic bands (still without current), the Taylor expansion of the
Lindhardt function in the low-energy low-frequency limit gives the following
result for the spin wave propagator:
 \begin{eqnarray}
D^{\rm ret}_{\pm}(\vec{q},\omega)
 =\frac{4U\Delta}{3}\frac{1}{\omega\pm\rho q^2} 
\label{prop}
\end{eqnarray}
where $\rho$ is  the spin stiffness which is easily computed analytically
for the case of parabolic bands.  The poles
of Eq.(~\ref{prop}) give the well known result for the spin wave dispersion, 
$\omega= \pm \rho q^2$. Several remarks are in order: 
{\em i)} 
In real systems, spin-orbit interactions lift spin rotational invariance, 
resulting in a gap for the $q=0$ spin waves.
The size of the gap is typically of order of 1 $\mu$eV \cite{skomskicoey}.
{\em ii)} The interplay between disorder and spin orbit interactions, absent in the
above model, gives rise to a broadening of the spin wave spectrum, even at
small frequency and momentum. In Section V we address this issue and discuss how 
damping is changed in the presence of a current. 

\subsection{Spin waves with current}

In the previous subsection we derived the spin wave spectrum of a metallic
ferromagnet in thermal equilibrium. Equations (\ref{prop0}) and
(\ref{Linhardt}) establish a clear connection between spin waves and
quasiparticle distributions. In order to address the same problem in the
presence of a current, a non-equilibrium formalism is needed. 
By taking advantage of the formulation discussed above in which 
collective excitations interact with fermion particle-hole excitations we are 
able to appeal to established results for harmonic oscillators weakly
coupled to a bath.  In the equilibrium case, the fact that the low-energy 
Hamiltonian for magnetization-orientation fluctuations is that of a 
harmonic oscillator follows by expanding the fluctuation action to leading
order in $\omega$ to show that $\hat y$ and $\hat z$ direction fluctuations
are canonically conjugate.  In our model  
magnons are coupled to a bath of spin-flip particle-hole excitations.
Following  system-bath weak coupling master equation analyses\cite{CT} 
we find that the collective dynamics in the presence of a non-equilibrium
current-carrying quasiparticle system differs from the equilibrium one 
simply by replacing Fermi occupation numbers by the non-equilibrium
occupation numbers of the current-carrying state.
The following term therefore appears in the Taylor expansion of
the Lindhardt function $\Gamma$:
\begin{equation}
\left. \frac{\partial \Gamma}{\partial
q_i}\right|_{q=\omega=0}=
\frac{1}{{\cal N}\Delta^2} 
\sum_{\vec{k}} \frac{\partial \epsilon(\vec{k})}{\partial k_i}
 \left[
n^{\uparrow}_{\vec k}-n^{\downarrow}_{\vec k}   \right]
\end{equation}
Since this expression uses the easy direction $\hat x$ as the spin-quantization axis, 
the $x$ (spin) component of the {\em spin current} is:
\begin{equation}
\vec{\cal J} \equiv
\frac{e}{{\hbar \cal N}}
\sum_{\vec{k}}
 \frac{\partial \epsilon(\vec{k})}{\partial \vec k}
  \left[
n^{\uparrow}_{\vec k}-n^{\downarrow}_{\vec k}   \right]
\end{equation}
so that 
\begin{equation}
\left. \frac{\partial \Gamma}{\partial
q_i}\right|_{q=\omega=0}=
\frac{\hbar }{e\Delta^2} 
{\cal J}_i
\label{spincurrent}
\end{equation}
The   quantity ${\cal J}_i$,  the component of the spin current polarized along the 
magnetization direction ${\bf n}=\hat x$ and flowing 
along the $i$ axis, is the difference between the current carried by majority
and minority carriers. In equilibrium there is no current and no linear term  
occurs in the wavevector Taylor series expansion, leading to quadratic magnon 
dispersion as obtained in Eq.(~\ref{prop}).  When (charge) current flows
through the ferromagnet, the difference in carrier density and mobility between
majority and minority bands inevitably gives rise to a nonzero spin current
 \cite{Fert}.
We therefore obtain the following spectrum for spin waves in the presence of a current:
\begin{equation}
\omega=\rho q^2 - \frac{2U}{3\Delta}\frac{\hbar}{e}
\vec{q}\cdot\vec{\cal J} 
\label{main0}
\end{equation}
This equation is the central result of our paper.  Notice that it is in precise
agreement with the single-mode-approximation expression since 
$\Delta = \frac{2U}{3} (n_{\uparrow} - n_{\downarrow})$; in that case, however,
the explicit expression was derived for the case of free-particle parabolic bands only.  
Eq.(~\ref{main0}) states that
the spin wave spectrum of metallic ferromagnet driven by a current is modified
in proportion to the resulting spin current.

In the half metallic case, when the density of minority carriers is zero, the
spin current is equal to the total current and we recover the result of BJZ
\cite{BJZ}. In that limit $\Delta = \frac{2U}{3}n$ 
and $\rho\simeq \frac{\hbar^2}{2m}$,
leading to 
\begin{equation}
\omega=\frac{\hbar^2}{2m}q^2 - \frac{\hbar}{e n} \vec{q} \cdot \vec{j} =
\frac{\hbar^2}{2m} q^2 - \hbar \vec{q}\cdot\vec{v}_D
\label{main1}
\end{equation}
where we have expressed the current as $\vec{j}= e n \vec{v}_D$ with 
$\vec{v}_D$ the {\em drift} velocity, generalizing the half-metallic
simple Doppler shift result to non-parabolic bands.

\subsection{Spin wave instability}

Eqs. (~\ref{main0}) and (~\ref{main1}), taken at face value, predict that the energy of
a spin waves is negative and therefore that the uniform ferromagnetic state is 
destabilized by an arbitrarily small current.  If this were really true, it 
would presumably be a rather obvious and well known experimental fact.
It is not true because spin waves in real ferromagnetic materials have a gap due to both spin-orbit
interactions and magnetostatic energy.   Inserting by hand this (ferromagnetic resonance)
gap, the spin wave dispersion reads:
\begin{equation}
\omega=\omega_0+ \rho q^2 - \frac{2U}{3\Delta}\frac{\hbar}{e}
\vec{q}\cdot\vec{\cal J} 
\label{main2}
\end{equation}
so that it takes a {\em critical} spin current to close the spin wave gap.
In Fig.(2) we plot the current driven spin wave spectrum assuming $\omega_0=1
\mu eV$, the electronic density of iron ($n=1.17\; 10^{23}\;cm^{-3}$) and a
Doppler shift given by $q\;v_D$.  The critical current so estimated is
$\sim 1.1\;10^9$ A cm$^{-2}$ for a typical system. 
This critical current could be much lower, perhaps by several orders 
of magnitude, in metallic ferromagnets in which material parameters have been
tuned to minimize the spin-wave gap.  Experimental searches for current-driven 
anomalies in permalloy thin films, for example, could prove to be fruitful.

\begin{figure}
\includegraphics[width=2.6in]{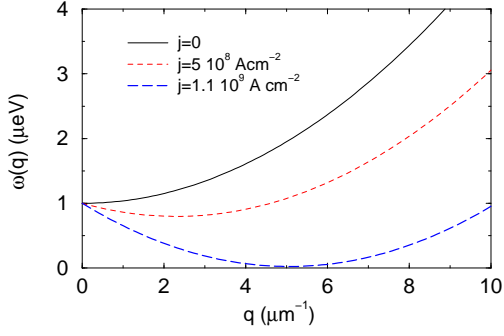}
\caption{ \label{fig2} Current modified spin-wave spectrum}
\end{figure}

\subsection{ Spin wave action with current}

In the small $\omega$ and small $\vec{q}$ limit, the spin waves are 
independent and their action is equivalent to that of an ensemble of non interacting harmonic
oscillators,indexed with the label $\vec{q}$. 
The Matsubara action for a single oscillator mode is the frequency sum of
\begin{eqnarray}
\left[p_{\vec{q}},x_{\vec{q}} \right]
\left( \begin{array}{cc}
\frac{1}{2M_{\vec{q}}} & -i\frac{\omega}{2}
\\
i \frac{\omega}{2} & \frac{K_{\vec{q}}}{2}
\end{array}
\right)
\left[
\begin{array}{c}
p_{\vec{q}} \\ x_{\vec{q}}
\end{array}
\right]
\end{eqnarray}
where the diagonal terms are the Hamiltonian part of the action and the
off-diagonal term can be interpreted as a Berry phase. For the
spin waves, the analog of $p$ and $x$ are, modulo some constants, 
 $\delta\Delta_y,\delta\Delta_z$. In this representation,
 the low $\omega$ and low $\vec{q}$ spin wave action reads:
\begin{eqnarray}
{\chi}_{\perp}^{-1}(\omega,\vec{q})=
\left( \begin{array}{cc}
\rho \vec{q}\cdot\vec{q} & 
-i\omega
\\
i\omega& \rho
\vec{q}\cdot\vec{q}
\end{array}\right)
+\frac{2U}{3 \Delta} \frac{\hbar}{e}
 \vec{\cal J}\cdot \vec{q}
 \left( \begin{array}{cc}
0 & 
-i
\\
i& 0
\end{array}\right)
\label{SWyz}
\end{eqnarray}
This representation makes it clear that the spin wave Doppler shift appears as
a modification of the term which couples the canonically conjugate variables,
$\delta \Delta_y$ and $\delta \Delta_z$, {\em i.e.}, the spin wave Doppler
shift modifies the {\em Berry phase}.  When expressed in this way, the spin-wave
Doppler shift is partly analogous to the change in superfluid velocity in a 
superfluid that carries a finite mass current, and the stability limit we 
have discussed is partly analogous to the Landau criterion for the critical 
velocity of a superfluid.  These analogies are closer in the case of ideal 
easy-plane ferromagnets, which like superfluids have collective modes with 
linear dispersion instead of a having a gap.  

\section{Alternate derivation of spin-wave Doppler shift}

In the previous section we have used a functional integral
approach to calculate how the spin wave {\em propagator} of a Hubbard model
metallic ferromagnet is modified when current flows through the
system.  The BJZ derivation of the same effect was based on an identity at the operator
level.  BJZ used a $s-d$ model, {\em i.e.} a Hamiltonian for itinerant ($s$)
electrons interacting with localized ($d$) spins via an exchange interaction.
They considered the limit of very large exchange interaction and low $s$
electron density, so that, in the ground state, the electrons are fully
spin polarized. They then introduced a local spin rotation transformation defined 
so that {\em at
every point of the space} the spin of the $s$ electrons is parallel with the
local value of the $d$ electrons magnetic moment. This unitary transformation
has been previously used for both  $s-d$ and other microscopic models of ferromagnetism 
\cite{Schulz,Prange1,Millis,Levy,Tatara}. In the transformed frame, the
exchange interaction is always diagonal in the spin index but the expression
for the kinetic energy is complicated, and includes new terms. One of the new terms 
couples the $s$ electron current to a space derivative of the local spin
magnetization.  It is from this term in the exchange energy that 
BJZ derived the modification of the Landau-Lifshitz equations that we have
identified as a spin-wave Doppler shift. 
In this section we bridge the gap between the two derivations.  We
recover the half metallic s-d Hamiltonian result of BJZ in a systematic way.

The continuum s-d model describes itinerant electrons,
$\psi_{\sigma}$, interacting with a continuum of localized quantum spins,
$\vec{M}(\vec x)$, through a exchange interaction of strength $J$. The
Hamiltonian for parabolic bands  is given by:
\begin{eqnarray}
{\cal H}=\int d\vec{x}\sum_{\sigma,\sigma'} \psi^{\dagger}
_{\sigma}\left( -\frac{\hbar^2 \vec{\nabla}^2}{2 m}
\delta_{\sigma,\sigma'} - 
\frac{J\vec{\tau}_{\sigma,\sigma'}}{2}\cdot \vec{M}(\vec{x})\right)
\psi_{\sigma'} 
\nonumber
\end{eqnarray}
where $\vec{\tau}$ are the  Pauli matrices.  In order to derive an effective
theory for the collective behavior of this system, we express its partition
function as a coherent state path integral: 
$${\cal Z}=\int {\cal D}^2{\Psi}_{\sigma}(\vec{x},\tau)
{\cal D}\vec{\Omega}(\tau) 
e^{-S_B + \int_0^{\beta}d\tau 
\overline{\Psi}_{\sigma'}(\partial_{\tau}-\mu)
\Psi_{\sigma'} - {\cal H}} $$
where $\tau$ is imaginary time, $\vec{\Omega}(\vec{x},\tau)=\frac{1}{S}\vec{M}(\vec{x},\tau)=
\left[\cos(\phi)\sin(\theta),\sin(\phi)\sin(\theta),\cos(\theta)\right]$ is the 
unimodular vector field which labels the spin coherent states,  $S_B$ is the
Berry phase term that captures the spin commutation relations \cite{Auerbach}, and $\Psi$ are the Grassmann
numbers which label the fermion coherent states \cite{Negele}.

Following BJZ, we perform a unitary transformation on the spins of the
itinerant electrons so that, at each point of time and space, the quantization
axis is parallel to  $\vec{M}(\vec x,\tau)$.  BJZ considered only the limit of
very strong  ferromagnetic $J$, so that the spins of the occupied electronic
states are always parallel to $\vec{M}(\vec{x},\tau)$ and we can ignore the
antiparallel electrons. This approximation is valid in half-metallic systems
for energies much smaller than $J$, the local spin splitting. In this
approximation the  action for the parallel fermions in the rotated frame,
denoted by $\Phi(\vec{x},\tau)$,  can then be written as $S=S_B+ S_0 +S_1 +S_2$ where:
\begin{eqnarray}
&S_B&=\int_0^{\beta} d\tau\int d\vec{x}  \;i\;S\;c_S
\left( 1 + 
\frac{\overline{\Phi}{\Phi}}{2Sc_S}\right) \cos(\theta)
\partial_{\tau}\phi
\nonumber \\
&S_0&= \int_0^{\beta}d\tau \int d\vec{x}\;
  \overline{\Phi} \left(\partial_{\tau} -\mu -\frac{\hbar^2 \nabla^2}{2 m} -
   JS \right) \Phi \nonumber \\
&S_1&= \frac{\hbar^2}{2m}\int_0^{\beta}d\tau\int 
d\vec{x} \left(\sum_{i,j} \nabla_i \Omega_j \;\; \nabla_i \Omega_j
\right)  \frac{1}{4} \overline{\Phi}{\Phi} 
\nonumber \\
&S_2&= \frac{2}{ec} \int_0^{\beta}d\tau\int d\vec{x} \left[ \vec{\cal J}_{P}
+ \vec{\cal J}_D \right]\cdot \vec{A}(\hat{\Omega}),
\label{action3}
\end{eqnarray}
$c_S$ is the density of local moments with spin $S$, and  
$\vec{A}(\hat{\Omega})=\frac{\hbar c}{4}\cos(\theta)\vec{\nabla}\phi$ 
is an {\em effective} vector potential which depends on the local spin
configuration, $\vec{\Omega}$.
In Eq.~\ref{action3}, $\vec{\cal J}_{P}$ and $\vec{\cal J}_{D}$ are respectively
{\em paramagnetic} and {\em diamagnetic} contributions to the current density defined by 
\begin{eqnarray}
\vec{\cal J}_{P}&\equiv& \frac{e \hbar}{2 m i}  \left(
\overline{\Phi}(\vec{x},\tau) \vec{\nabla}{\Phi}(\vec{x},\tau)
-\vec{\nabla}\overline{\Phi}(\vec{x},\tau){\Phi}(\vec{x},\tau) 
\right)
\nonumber \\
\vec{\cal J}_{D}&\equiv&
\frac{e}{mc}  \overline{\Phi}(\vec{x},\tau){\Phi}(\vec{x},\tau)
 \vec{A}(\hat{\Omega}).
\end{eqnarray}

The $\vec{\cal J}_P\cdot\vec{A}$ coupling has the form anticipated by 
BJZ.  To address the magnetic elementary excitation spectrum we formally integrate out the 
fermion fields $\Phi$ and expand to quadratic order in magnetic fluctuations.  The 
action expressed in terms of only the spin fields is
$S_{\rm eff}(\vec{\Omega})= S_B + {\rm Tr}[ \ln {\cal G}^{-1}]$ with 
\begin{eqnarray}
 {\cal G}^{-1}(\theta,\phi)=\partial_{\tau} -\mu - \frac{\hbar^2}{2m}
\left(i\vec{\nabla} -\frac{\vec{A}}{2c} \right)^2  + 
\nonumber \\  +i \cos(\theta)
\partial_{\tau}\phi + \
 \frac{\hbar^2}{8m}\sum_{i,j} \nabla_i \Omega_j \; \nabla_i \Omega_j.
 \end{eqnarray}
Expanding around the $\hat x$ ($\theta=\frac{\pi}{2}\; \phi=0$) direction
we obtain for the spin-wave action 
\begin{eqnarray}
{\rm TrLn}\left[{\cal
G}^{-1}(\theta,\phi)\right]={\rm TrLn}\left[{\cal
G}^{-1}(\frac{\pi}{2},0)+ \delta{\cal G}^{-1}(\vec{\Omega})\right] 
\nonumber 
\end{eqnarray}
To leading order in $\delta{\cal G}^{-1}$, the action reads:
\begin{eqnarray}
&S&=S_B(n)
+ \int_0^{\beta}\int d\vec{x} 
\frac{\hbar^2 n}{8m}\sum_{i,j} \nabla_i \Omega_j \;\; \nabla_i \Omega_j
 \nonumber \\
&+& \frac{2}{ec} \int_0^{\beta}\int d\vec{x}  \left[ \vec{j} 
+\frac{e n}{mc}  \vec{A}(\hat{\Omega}) \right]\cdot \vec{A}(\hat{\Omega})
\label{action5}
\end{eqnarray}
where $\vec{j}\equiv  Tr\left[{\cal G}(0,\frac{\pi}{2}) \vec{\cal
J}_{P}(\vec{x},\tau) \right]$ is the average current and  $n\equiv
Tr\left[{\cal
G}(0,\frac{\pi}{2})\overline{\Phi}(\vec{x},\tau){\Phi}(\vec{x},\tau)\right]$ 
is the  average density in the collinear ground state.  In deriving this
expression we allowed the mean-field fermion quasiparticle occupation
numbers to assume values consistent with a non-equilibrium current-carrying
state.

Equation (\ref{action5}) defines a theory for the collective magnetization of
the ferromagnet. The first two terms are the Berry phase of the $d$ spin and  a
renormalization of the Berry phase due to the spin of the $s$ electrons,
similar to that derived by Millis {\em et al.} for the double exchange model
\cite{Millis}. The third term describes the energy penalty for non collinear
configurations, or spin stiffness. The terms in the second line yield the
coupling of the average (paramagnetic and diamagnetic)  currents to the
collective magnetization.  

The semiclassical equations of motion of (\ref{action5}) yield the Landau
Lifshitz (LL) equations including the $j_i \nabla_i \Omega \times \Omega$  term
derived by BJZ (equation (5)).  In the case of BJZ, the LL equations are derived from a 
micromagnetic energy functional plus the paramagnetic current term. In our case, the
whole functional is derived from the microscopic Hamiltonian.   
The spin wave expansion for (\ref{action5}) around a 
classical homogeneous ground state, $\vec{\Omega}_{\rm cl}= \hat x$ 
is obtained by  expanding
 $\vec{\Omega}=\vec{\Omega}_{\rm cl} + \delta{\vec{\Omega}}$ and identifying
 $\delta\;\Omega_y \simeq \phi$, $\delta \;\Omega_z\simeq cos(\theta)$. 
Dropping terms of order $\delta\Omega^3$ and higher, the action
 (\ref{action5}) becomes:
\begin{equation}
 S_{SW}= \frac{1}{2\beta V}\sum_{{\cal Q},ab} 
\delta\Omega_a({\cal Q}) {\cal K}_{ab}({\cal Q})
\delta\Omega_b(-{\cal Q})
\label{SW2}
\end{equation}
as in Eq.(~\ref{spinfluc}).  After analytical continuation, the 
spin wave kernel 1, in the $y,z$ representation: 
\begin{eqnarray}
{\chi}_{\perp}^{-1}
&=&c_S
\left( \begin{array}{cc}
\rho  q^2 & -iS'\omega
\\
i S' \omega & \rho  q^2 
\end{array}\right)
+
\frac{\hbar}{e}
 \vec{ j}\cdot \vec{q}
 \left( \begin{array}{cc}
0 & 
-i
\\
i& 0
\end{array}\right)
\label{SWyz2}
\end{eqnarray}
where $r\equiv\frac{n}{c_S}$, $\rho\equiv r\; \frac{\hbar^2}{4m}$, 
and $S'=S+\frac{r}{2}$.   The main difference between $s-d$ and Hubbard model 
result is the appearance here of both local moment and itinerant electron
($r/2$) contributions to the Berry phase, which is proportional to the total spin density.
Note that since $\vec{A}$ is quadratic in the spin wave
variables, the term $\vec{A}^2$ in (\ref{action5}) gives no contribution to
(\ref{SW2}). 
After diagonalization of Eq. (\ref{SWyz2}) we obtain the retarded propagator
for the spin wave variables. The real and imaginary part of the poles of the
retarded propagator give the spin wave dispersion and damping, respectively. 
In this theory, the  imaginary part is zero, since the spin flip of
quasiparticles is blocked.    The real part reads: 
\begin{equation}
\omega=\left[\frac{\hbar^2 n \vec{q}^2}{4S m }-\frac{\hbar}{2Se} \vec{j}
\cdot\vec{q} \right]
\times\frac{1}{c_S\left(1+ \frac{n}{2Sc_S}\right)}
\label{main3}
\end{equation}

Hence, we see how the spin wave dispersion in this theory has the
 $\vec{q}\cdot\vec{j}$ term derived by BJZ. Since the system described by the
 theory is fully polarized, the current and the spin current (polarized along the
 ground state magnetization direction) are identical.  This result is to be compared with
 Eq.~(\ref{main1}), derived with a different method for a different
 microscopic model.  We conclude that spin-wave Doppler shifts due to spin currents are 
generic, although their quantitative details can depend on the microscopic physics of the 
ferromagnet. 

 
\section{Enhanced Spin-Wave Damping at finite Current}

In Sections III and IV we have shown how the dispersion of spin waves in
a metallic ferromagnet is affected by current flow, and we have obtained
results compatible with those of BJZ \cite{BJZ}. In this section we address a
problem which, to our knowledge, has remained unexplored so far: how does the
current flow affect the lifetime of the spin waves.  In  subsection A we
analyze the damping of spin waves at zero current. In the subsections B and
C we discuss how these results are modified by the presence of a
current. 

A ferromagnetic resonance (FMR) experiment probes the dynamics of the
coherent or $\vec{q}=0$ spin wave mode.  The signal linewidth is 
inversely proportional to the coherent mode lifetime, the time that 
it takes for a transverse magnetic fluctuation to relax back to 
zero. Spin waves have a finite lifetime because they are coupled to each other and 
to other degrees of freedom, including phonons and electronic quasiparticles.
In ferromagnetic metals, the quasiparticles are an important part of the 
dissipative environment of the spin waves
\cite{Kambersky,dampingref,Prange2,Bergerdamping}.
and we can therefore expect that quasiparticle current flow affects the
spin wave lifetime to some degree.  In order to discuss this effect, it 
is useful to first develop the theory of quasiparticle spin-wave damping in 
equilibrium.  

\subsection{Damping at zero current}

The elementary excitation energies for the ferromagnetic phase of the Hubbard
model, are specified by the locations of poles in Eq.(~\ref{prop0}).  The
damping rate is proportional to the imaginary part of the transverse
fluctuation propagator.  According to Eq.(~\ref{prop0}), the damping of a spin
wave with frequency $\omega$ and momentum $\vec{q}$, $\gamma(\vec{q},\omega)=
-2 {\rm Im} \left[ \Gamma(\omega,\vec{q}) \right] $ is given by: 
\begin{equation}
\gamma(\vec{q},\omega)= 
\frac{2\pi}{\cal N} \sum_{\vec{k}}
\left[n^{\uparrow}_{\vec{k}}-n^{\downarrow}_{\vec{k}+\vec{q}}\right]
\delta\left[ \epsilon^{\uparrow}_{\vec{k}}-\epsilon^{\downarrow}_{\vec{k}+\vec{q}}
 + \omega \right]
\label{damping0}
\end{equation}
In the absence of disorder, this quantity is nonzero when $|\vec{q}|$ is
comparable to  $k_{F\uparrow}-k_{F\downarrow}$ or when $\omega \simeq \Delta$,
the band spin-splitting. Either disorder, which breaks translational symmetry
leading to violations of momentum conservation  selection rules, or spin-orbit
interactions, which cause all quasiparticles to have mixed spin character, will
lead to a finite electronic damping rate at characteristic collective motion
frequencies.   Because this damping is extrinsic, however, its numerical value
is usually difficult to estimate.  It is often not known whether coupling to 
electronic quasiparticles, phonons, or other degrees of freedom dominates the 
damping.  

Formally generalizing Eq.(~\ref{damping0}) to the case with disorder and
spin orbit interactions leads to 
\begin{equation}
\gamma(\omega)\propto \sum_{\vec{k},\vec{k}',\nu,\nu'}
S_{\nu,\nu'}(\vec{k},\vec{k}')
\left(n^{\nu}_{\vec{k}}-n^{\nu'}_{\vec{k}'}\right)
\delta\left[ \epsilon^{\nu}_{\vec{k}}-\epsilon^{\nu'}_{\vec{k}'}
 + \omega \right]
\label{damping1}
\end{equation}
where $
S_{\nu,\nu'}(\vec{k},\vec{k}') \equiv
 |\langle \vec{k},\nu
|S^{(-)}|\vec{k'},\nu'\rangle|^2 $
is a matrix element between disorder broadened initial and final quasiparticle
states, labeled by momentum $\vec{k}$ and band index $\nu$ (but not Bloch
states).  Averaging out the extrinsic dependence on wavevector labels by letting
$S_{\nu,\nu'}(\vec{k},\vec{k}') \to S_{\nu,\nu'}$
we obtain 
\begin{eqnarray}
\gamma(\omega) &=&
n^2 \sum_{\nu,\nu'}
S_{\nu,\nu'} 
\int d\epsilon \int d\epsilon'
N_{\nu}(\epsilon)N_{\nu'}(\epsilon') 
\times \nonumber \\ &\times&
\left(n(\epsilon)-n(\epsilon')\right)
\delta\left[ \epsilon-\epsilon' + \omega \right]
\label{damping2}
\end{eqnarray}
where $N_{\nu}(\epsilon)$ is the density of states of the band $\nu$. 
For  $\omega$ of the order of the ferromagnetic resonance frequency, we can
expand Eq. (\ref{damping2}) to lowest order in $\omega$:
\begin{eqnarray}
\gamma(\omega)\simeq
\omega \left[ n^2 \sum_{\nu,\nu'}
S_{\nu,\nu'} 
N_{\nu}(\epsilon_F)N_{\nu'}(\epsilon_F) \right]
\label{damping3}
\end{eqnarray}
This result can be considered a microscopic justification of the Gilbert
damping law, which states that the damping rate is linearly proportional to the
resonance frequency and vanishes at $\omega=0$.  The proportionality between 
frequency and damping rate follows from 
phase space considerations: the higher the spin wave frequency $\omega$, the
larger the number of quasiparticle spin flip processes compatible with energy
conservation. 

\subsection{Damping at finite current}

We analyze how a current modifies quasiparticle damping, we again appeal to the
picture of magnons as harmonic oscillators coupled to a bath of  particle-hole
excitations and borrow results from master equation results  for oscillators
weakly coupled to a bath \cite{CT} For magnetization in the `$\uparrow$'
direction, magnon creation is accompanied by quasiparticle-spin raising and
magnon annihilation is accompanied by  quasiparticle-spin lowering.  It turns
out \cite{CT} that only the difference between the rate of quasiparticle
up-to-down and  quasiparticle down-to-up transitions enters the equation that
describes  the magnetization evolution.  This transition rate difference leads
to  the same combination of quasiparticle occupation numbers as in
Eq.(~\ref{damping3}), except that the occupation numbers characterize the
current-carrying state and  are not Fermi factors.  For metals we can use the
standard approximate form\cite{Ashcroft}  for the quasiparticle distribution
function in a current carrying state:
\begin{equation}
g^{\nu}_{\vec{k}}=n^{\nu}_{\vec{k}}-e\vec{E}\cdot \vec{v}_{\nu}(\vec{k})
\tau_{\nu}(\epsilon^{\nu}_{\vec{k}})
\left[-\left.\frac{\partial n}{\partial\epsilon}\right|
_{\epsilon=\epsilon^{\nu}_{\vec{k}}}\right]
\label{rta}
\end{equation}
Because of the independent sums over $\vec{k}$ and $\vec{k'}$ in 
Eq.(~\ref{damping1}), and because it is a simple difference of Fermi factors
that enters the damping expression, we conclude that the quasiparticle damping 
correction will vanish to leading order in the spin-dependent 
drift velocities $v^{\sigma}_D$.  We reach this conclusion even though the 
phase space for spin-flip quasiparticle transitions at the spin-wave energy
is altered by a factor $\sim 1$ when $\epsilon_F\times\frac{v_D}{v_F} \sim \epsilon_0$,
where $\epsilon_F$ is a characteristic quasiparticle energy scale,
{\em i.e.} the up-to-down and down-to-up transition rates change significantly
when this condition is met, but not their difference.  
To obtain a crude estimate for the current at which this condition
is satisfied we use the 
following data\cite{Ashcroft} for iron: $n\approx$ 1.7 10$^{23}$, Fermi
velocity $\sim 1.98$ 10$^{8}$ cm s$^{-1}$.   The drift velocity corresponding
to  a current density of $10^{\beta}A$ cm$^{-2}$ is $v_d= \frac{j}{en} \simeq 
10^{\beta-4}$ cm s$^{-1}$.  The typical energy of a long-wavelength magnon is
$\sim 10^{-6}$ eV. Therefore, current densities of the order of $10^6$
A cm$^{-2}$ and larger will substantially change the coupling of spin-waves
to their quasiparticle environment.  Although this change will influence 
the spin-wave density-matrix, magnetization fluctuation damping itself will not be
altered by this mechanism until much stronger currents are reached.  

\subsection{Two magnon damping}

In the previous subsections we have calculated the damping of the lowest energy
spin wave due to its coupling to the reservoir of quasiparticles. In this
section we study damping of the coherent rotation mode ($\vec{q}=0$ spin
wave) due to its coupling to finite $\vec{q}$ spin waves. This mechanism
is known as {\em two magnon scattering} and is efficient when the coherent
rotation mode is degenerate with finite $\vec{q}$ spin waves \cite{Mills}, a 
circumstance that sometimes arises due to magnetostatic interactions. The main
point we wish to raise here is that because of the spin-wave Doppler shift, 
precisely this situation arises when the ferromagnet is driven by a current.
As in the previous subsection, we assume that some type of
disorder lifts momentum conservation. The effective Hamiltonian for the
spin waves reads:
\begin{equation}
H=\omega_0 b^{\dagger}_0 b_0 + \sum_{\vec{q}\neq 0} \omega(\vec{q})
b^{\dagger}_{\vec{q}} b_{\vec{q}} + b^{\dagger}_0\sum_{\vec{q}\neq 0}
\frac{g_{\vec q}}{\sqrt{V}}
b_{\vec q} + {\rm h.c.}
\label{hamilbos}
\end{equation}
where $b_{\vec q}$ is the annihilation operator for the spin wave with momentum
$\vec{q}$ and $g_{\vec{q}}$ is some unspecified matrix element accounting for
disorder induced elastic scattering of the spin waves. 
Equation (\ref{hamilbos})  is the well Hamiltonian known for a 
damped harmonic oscillator and can be solved exactly.
 The damping rate for the $\vec{q}=0$ spin wave reads:
\begin{equation}
\gamma(\vec{\cal J})= \frac{2 \pi}{\hbar} \int \frac{d \vec{q}}{(2
\pi)^3} |g_{\vec{q}}|^2 \delta(\omega_0-\omega_{\vec{q}})
\end{equation}
Now we use $\omega_0-\omega_{\vec{q}}=\rho q^2 -a\vec{q}\cdot\vec{\cal J}$. 
After a straightforward calculation we obtain:
\begin{equation}
\gamma(\vec{\cal J})= \frac{g^2}{4 \pi} \frac{ a|\vec{\cal J}|}{\rho^2}
\label{damping2sw}
\end{equation}
where we have approximated $g_{\vec{q}}\simeq g$. Hence, in the presence of
elastic spin wave scattering, renormalization of the spin wave spectrum due
to the current will enhance the damping of the lowest spin wave mode. Unlike the 
Gilbert model, the damping rate given by
equation (\ref{damping2sw}) is independent of $\omega_0$, implying that the 
dimensionless Gilbert damping coefficient would decline with external field
if this mechanism were dominant. 

\section{Spin-wave Doppler shift as a Spin-Torque Effect} 

In this section we explain how the influence of an uniform current on
magnetization dynamics can be understood   as a special case of a spin-torque
effect\cite{Slon,Berger}. The latter takes place when a spin current coming
from a magnet spin polarized along $\vec{M}_1$ enters in a  magnet spin
polarized along $\vec{M}_2$.  In this circumstance
 there is an imbalance between the
incoming and the outgoing transverse component (with respect to $\vec{M}_2$) of
the  spin currents in magnet 2. Because of spin conservation (resulting from
the rotational invariance of the system), the imbalance in the spin flux across
the boundaries of magnet 2   {\em must be compensated} by a change of the
magnetization of that magnet, which is described by a new term in the Landau
Lifshitz equation   \cite{Slon,Berger}. The microscopic origin of the spin
current imbalance can be understood as a destructive interference effect,
originated by the fact that the steady state spin current is a sum over
stationary states with broad distribution in momentum space \cite{Slon}.
Alternatively, it is possible to understand the spin current flux imbalance as
a destructive interference in the time domain. At a given instant of time, the 
outgoing current-carrying quasiparticles have elapsed a different amount of
time in magnet 2. This broadening in the interaction time distribution results
in a broadening of the spin precession angle \cite{us-condmat}. The average
over that distribution results in a  vanishing  transverse spin component of
the  outgoing flux. 

The above argument, connecting spin flux imbalance and spin-torque, applies to
a system in which the inhomogeneous magnetization is described by  piecewise
constant function.  It is our contention that the spin wave Doppler shift can
be understood by applying the same argument to the case  of smoothly varying
magnetization.  We consider again a  magnet with charge current  $\vec j$, and
spin current  $\vec{\cal J}$. We assume that current flows in the $\hat x$
direction and, importantly,  that the spin current is locally parallel to the
magnetization orientation $\vec{\cal J}(x)=j_s\hat{\Omega}(x)$.  It can be
shown that this is the case in a wide range of situations. 

The spin density reads $\vec{S}(x)= S_0 \hat{\Omega}(x)$ where $S_0$ is the
average spin polarization. We focus on the slab centered at x and  bounded by
$x-dx$ and $x+dx$. Spins are injected into the slab at the rate $j_s
\hat{\Omega}(x-dx)$ and leave at the rate  $j_s\hat{\Omega}(x+dx)$. The
resulting spin current imbalance is $ 2 dx j_s \partial_x \hat{\Omega} $.
Therefore, there must be a spin transfer to the local magnetization:
\begin{equation}
\left.\frac{d\vec{S}(x)}{dt}\right|_{ST}
=j_s \partial_x \hat{\Omega} 
\end{equation}
Now, using $|\hat{\Omega}|^2=1$ at every point of the space we obtain:
\begin{equation}
\left.\frac{d\vec{S}(x)}{dt}\right|_{ST}
=j_s \hat{\Omega}(x)\times(\partial_x \hat{\Omega}(x)\times\hat{\Omega}(x)) 
\end{equation}
which is exactly the same result obtained in \onlinecite{BJZ}. Including this term
in the Landau Lifshitz equation and solving for small perturbations around the
homogeneous ground state (spin waves) results into the spin wave Doppler shift
discussed in previous sections. In conclusion, this argument demonstrates that
the spin-wave Doppler shift and spin transfer torques are different limits of
the same physical phenomena, the transfer of angular momentum from the
quasiparticles to the collective magnetization whenever the latter is not
spatially uniform.  

\section{Discussion and conclusions} 

The effect of high current densities on the magnetization dynamics of 
ferromagnetic metals have been explored experimentally in several
configurations. In point contact experiments, a large current density is
injected from a normal metallic contact into a ferromagnetic multilayer
\cite{Tsoi1,Tsoi2} or single layer \cite{Chien}. When a large flow of electrons ( 
current density j$\simeq 10^{8}$ A cm$^{-2}$)  enters  into the ferromagnetic
multilayer, the resistivity presents an abrupt increase which has been related
to the coherent precession of spin waves \cite{Tsoi1} and/or phonons
\cite{Tsoi2}. The fact that Ji and Chien \cite{Chien} report similar results when
the current is injected into a a single ferromagnetic layer demonstrates that 
interlayer coupling is not essential for the anomalies observed in transport. It
must be noted that when the current flow is such that the electrons go from
the ferromagnetic layer(s) toward the point contact, no anomaly is observed.  
Similar transport anomalies at currents densities higher than those of current
induced magnetization switching are observed by a number of different groups
 \cite{SMT-exp}
in a  system of two adjacent ferromagnetic nanopillars. In this system a large
current density  flows from one ferromagnet to the other. 

The fact that the current densities at which the anomalous behavior takes
place is of the same order  of magnitude than the current at which the spin
wave Doppler shift makes the collinear state unstable might lead to suggest a
connection between the two. However, the experiments in the point contact
geometry show that the transport anomalies only occur for one direction of
the current, something which seems at odds with the spin wave Doppler shift
instability. 

In summary, the focus of this paper is on the effect of the current in the spin
wave dynamics of a bulk ferromagnetic metal. We have addressed two types of
effects: the change in the spin wave dispersion and change in the spin wave
damping. These quantities are given, at a formal level, by the spin wave
propagator. The central idea is that the spin wave propagator is a functional
of the quasiparticle occupation function. In the presence of the current the
occupation function changes, affecting both the dispersion and the damping of
the spin waves. Throughout the paper we have assumed that the functional
relation between the quasiparticle occupation function and the spin wave
propagator remains the same when the system is out of equilibrium. In that
sense, the above derivations are heuristic.  Our main conclusions are: {\em i)}
A current $\vec{j}$ flowing through a metallic ferromagnet results\cite{Fert}
in a spin current  $\vec{\cal J}$ which modifies its  spin wave spectrum by an
amount proportional to $\vec{q}\cdot \vec{\cal J}$. {\em ii)} This
modification, which was derived by BJZ for a fully polarized s-d model, occurs
as well in a non fully polarized Hubbard model, in which the $d$ electrons are
itinerant and, according to the arguments of section II,  in  typical
real-world ferromagnets. {\em iii)} In the presence of elastic two  magnon
scattering,  the spin wave Doppler shift leads as well to a broadening of the
lowest spin wave mode  (Eq. \ref{damping2sw}), which is proportional to the
spin current.  {\em iv)} Both the spin-wave Doppler shift in spatially
homogeneous ferromagnets and the spin torque effect in inhomogenoeus structures
\cite{Slon,Berger} are a consequence of the spin transfer from the 
quasiparticles to the collective  magnetization when the latter is spatially
inhomogeneous. 

We  acknowledge fruitful discussions with M. Abolfath and M. Tsoi.
Work at the University of Texas was supported by the Welch
Foundation and by the National Science Foundation under grants
DMR-0210383 and DMR-0115947. Work at University of Alicante supported by Ram\'on y
Cajal Program, Ministerio de Ciencia y Tecnolog\'ia.

\widetext



\begin{references}


\bibitem{Maekawa} {\em Spin Dependent Transport in Magnetic Nanostructures}, Ed.
by S. Maekawa and T. Shinjo, (Taylor and Francis, 2002).

\bibitem{Slon} J.C. Slonczewski, J. Mag. Mat. Mag. {\bf 159}, L1 (1996).

\bibitem{Berger} L. Berger, Phys. Rev. B{\bf 54}, 9353 (1996).

\bibitem{Tsoi1} M. Tsoi {\em et al.}, Phys. Rev. Lett. {\bf 80}, 4281 (1998);
M. Tsoi et al., Nature {\bf 406}, 46 (2000); 

\bibitem{Tsoi2} M. Tsoi {\em et al.}, Phys. Rev. Lett. {\bf 89}, 246803 (2002).

\bibitem{Sun} J.Z. Sun, J. Magn. Mag. Mater. {\bf 202}, 157 (1999).

\bibitem{SMT-exp} E. B. Myers {\em et al.}, Science {\bf 285}, 867 (1999) 
J.A. Katine {\em et al.}, Phys. Rev. Lett. {\bf 84}, 4212 (2000); 
E.B. Myers, et al., Phys. Rev. Lett. {\bf 89}, 196801, (2002); 
S.I. Kiselev {em et al.}, preprint [cond-mat/0306259] (2003). 
W.H. Rippard, M.R. Pufall,and T.J. Silva,
 Appl. Phys. Lett. {\bf 82}, 1260-1262 (2003).
F. B.  Mancoff {\em, et al.}  Appl. Phys. Lett. {\bf 83}, 1596 (2003).


\bibitem{Chien} Y. Ji and C.L. Chien and M. D. Stiles, 
 Phys. Rev. Lett. {\bf 90}, 106601 (2003). 

\bibitem{MSU} S. Urazhdin et al., Phys. Rev. Lett. {\bf 91}, 146803 (2003)

\bibitem{TransferTheory} J. Z. Sun, Phys. Rev. B 62, 570 (2000 );
A. Brataas, Y. V. Nazarov, and G.E.W. Bauer, Phys. Rev. Lett. {\bf 84}, 2481 (2000);
X. Waintal and P.W. Brouwer, Phys. Rev. B {\bf 63}, 220407 (2001);
C. Heide, Phys. Rev. B {\bf 65}, 054401 (2002);
M. Stiles and A. Zangwill, Phys. Rev. B {\bf 65}, 014407 (2002);
J.-E. Wegrowe, Appl. Phys. Lett. {\bf 80}, 3775 (2002); 
S. Zhang, P.M. Levy, and A. Fert, Phys. Rev. Lett. {\bf 88}, 236601 (2002);
G.E.W. Bauer, Y. Tserkovnyak, D. Huertas-Hernando, and A. Brataas,
Phys. Rev. B {\bf 67}, 094421, (2003);
M.L. Polianski and P.W. Brouwer, preprint [cond-mat/0304069] (2003);
A. Shapiro, P. M. Levy, and S. Zhang Phys. Rev. B {\bf 67}, 104430 (2003); 
A. Fert {\em et al.},  preprint[cond-mat/0310737] (2003).

\bibitem{BJZ} 
Y.B. Bazaliy, B. A. Jones and S.-C. Zhang, Phys. Rev. B{\bf 57}, R3213
  (1998)

\bibitem{footnote} More precisely, 
$\vec{\cal J} \equiv \frac{e}{{\hbar \cal N}} \sum_{\vec{k}} \frac{\partial
\epsilon(\vec{k})}{\partial \vec k} \left[ n^{\uparrow}_{\vec
k}-n^{\downarrow}_{\vec k}   \right]$, 
where ${\cal N}$ is the number of sites in the lattice, and $\uparrow$ and
$\downarrow$ are defined in the axis of the average magnetization. 

\bibitem{Kittel} C. Herring and C. Kittel, Phys. Rev. {\bf 81}, 869 (1951).

\bibitem{bandpicture} E.P. Wolfarth, Rev. Mod. Phys. {\bf 25}, 211 (1953); D.A.
Papaconstantopoulos, Handbook of the Band Structure
of Elemental Solids (Plenum Press, New York, 1986). 

\bibitem{Prange4} R. E. Prange and V. Korenman, Phys. Rev. B {\bf 19}, 4691 (1979)

\bibitem{Moriya}  T. Moriya, {\em Spin Fluctuations in Itinerant Electron
Magnetism},  (Springer-Verlag, Berlin, 1985) 

\bibitem{Hubbard} J. Hubbard, Phys. Rev. B{\bf 19}, 2626 (1979)

\bibitem{Schulz} H.J. Schulz, Phys. Rev. Lett{\bf 65}, 2462 (1990)

\bibitem{Fradkin} Sections 3.3  and 3.4 in E. Fradkin,
{\em Field Theories of Condensed Matter Systems},
(Addison Wesley, Reading, MA, 1991)

\bibitem{Doniach} S. Doniach and E. H. Sondheimer, {\em Green's Functions for
Solid State Physicists}, W. A. Benjamin, Inc (1974)

\bibitem{HS}  R.L. Stratonovitch, Dok. Akad. Nauk. SSSR {\bf 115}, 1097 (1957);
 [Sov. Phys. Dokl. {\bf 2}, 416 (1958)]. 
 J. Hubbard, Phys. Rev. Lett. {\bf 3}, 77 (1959) 

\bibitem{Negele} J.W. Negele and H. Orland,
{\em Quantum Many-Particle Systems}, Addison Wesley, (1988)

\bibitem{Auerbach} Assa Auerbach,
{\em Interacting Electrons and Quantum Magnetism}, Springer-Verlag, 1994.

\bibitem{skomskicoey} R. Skomski and J. M. D. Coey, Permanent Magnetism
(Institute of Physics, Bristol, 1999).


\bibitem{CT}  Claude Cohen-Tannoudji, Gilbert Grynberg, Jacques Dupont-Roc 
{\em Atom-Photon Interactions: Basic Processes and Applications}, 
(Wiley-Interscience, New York,1998)  

\bibitem{Prange1} V. Korenman, J. L. Murray, and R. E. Prange
     Phys. Rev. B {\bf 16}, 4032 (1977) 

\bibitem{Millis} A. J. Millis, P.B. Littlewood and B. I. Shraiman, Phys. Rev.
Lett{\bf 74}, 5144 (1995)

\bibitem{Levy}  P. M. Levy and S. Zhang,    Phys. Rev. Lett. {\bf79} , 5110 (1997)

\bibitem{Tatara} G. Tatara and H. Fukuyama, Phys. Rev. Lett. 78, 3773 (1997) 

\bibitem{Fert} A. Fert, J. Phys. C. {\bf 2}, 1784 (1969) and references therein.


\bibitem{Kambersky}  V. Kambersky, Canadian Journal of Physics, {\bf 48} 2906 (1970)

\bibitem{dampingref}     V. Kambersky and C. E. Patton, Phys. Rev. B {\bf 11},
 2668 (1975)

\bibitem{Prange2} V. Korenman, J. L. Murray, and R. E. Prange
     Phys. Rev. B 16, 4048-4057 (1977)

\bibitem{Bergerdamping} L. Berger, J. Phys. Chem. Solids {\bf 38}, 1321 (1977)
 
\bibitem{Ashcroft} N. W. Ashcroft and N. D. Mermin, {\em Solid State Physics},
W.B. Saunders Company (1976)

\bibitem{Mills} R. Arias and D. Mills, Phys. Rev. B {\bf 60}, 7395 (1999)

\bibitem{us-condmat} J. Fernandez-Rossier {\em et al.}, 
 [cond-mat/0304492] (2003)


\end{references}
\end{document}